\documentclass[12pt]{article}

\title{On the Concept of Snowball Sampling}

\usepackage[round,longnamesfirst]{natbib}
\usepackage{csquotes,amsmath, amsthm, amssymb, amsbsy, mdwlist}

\usepackage[hyperfigures=true]{hyperref}

\author{Mark S. Handcock\footnote{Professor of Statistics, Department of Statistics, University of California, Los Angeles, CA 90095-1554 (E-mail: \emph{handcock@ucla.edu}).} \and
Krista J. Gile\footnote{Assistant Professor of Statistics,
  Department of Mathematics and Statistics,
  University of Massachusetts, Amherst,
  MA 01003-9305 (E-mail: \emph{gile@math.umass.edu}).}}

\begin{document}
\maketitle
Confusion over the definition of ``snowball sampling" reflects a phenomena
in the sociology of science: that multi-disciplinary fields tend
to produce a plethora of inconsistent terminology.  Often the
meaning of a term evolves over time, or different terms are used
for the same concept.  More confusing is the use of the same term
for different concepts.  The term
``snowball sampling" suffers from this treatment.

The term ``snowball sampling" has likely been in informal use for
a long time, but it certainly pre-dates \citet{coleman58} and \citet{trow57}.  
The earliest systematic work dates to the 1940s from the
Columbia Bureau of Applied Social Research, lead by Paul
Lazarsfeld.  The Bureau became interested in the empirical study
of personal influence via media \citep{barton2001}.  This led to the
consideration of interpersonal environments and to the
identification of opinion leaders and followers.  However
standard sampling of individuals was regarded as ineffective in
studying the relations between opinion leaders and followers as
pairs related in this way were seldom both selected in the sample
\citep[pp. 49-50]{lazarsfeld1944}.  To address this, Robert
Merton asked individuals in an initial diverse sample to name the
people who influenced them.  From these, a second wave of
influential people were interviewed as a ``snowball sample"
\citep{merton49}.  This approach was expanded in a panel survey of
women in a Midwestern town in 1945 \citep{katz.lazarsfeld:bk:1955}.
\citet{barton2001} provides a history of the work of the Bureau that
is still relevant to today's study of social media.

Trow's objective was to understand the support for
anti-democratic popular movements.  To do this he conducted an
empirical study of the political orientations and behaviors of
men in Bennington, Vermont in 1954 with particular focus on their
support for Senator McCarthy.  Trow conducted a snowball sample
over the friendship networks of the men starting from
``arbitrarily chosen lists of employees and occupational 
groups." \citep[p.  297]{trow57}.  He is very clear that this does not
produce a representative sample, and goes on to provide a
discussion of the issues with network sampling that is still
relevant today \citep[pp.  290-295]{trow57}.  He surmises: ``The
resulting sample, while not meant to be representative of any
specific population, nevertheless includes representatives of all
the important occupational groups, ..."

Following on from these foundations, 
\citet{coleman.et.al:soc:1957}
used the approach to collect information on influence
patterns among physicians.  
\citet{coleman58} is now the primary
reference for the meaning of snowball sampling.  He defines it
as: ``Snowball sampling: One method of interviewing a man's
immediate social environment is to use the sociometric questions
in the interview for sampling purposes." and describes Trow's
work as the example.

Acknowledging \citet{coleman58}, \citet{goodman:ams:1961} introduced ``s stage
k name snowball sampling", a specific form of snowball sampling.
Goodman's formulation requires an initial sample drawn using a
probability method on a known sampling frame.  It also fixes
parameters of the sampling process: the number of links followed
from each participant (k) and the number of waves of the sample
(s).  In this work, Goodman develops a rigorous statistical
approach to estimating certain relational features (number of
mutual ties, triangles, etc.) based on the resulting sample.
Just as \citet{lazarsfeld1944} followed links because they were
interested in studying, and therefore sampling, relationships
rather than individuals, Goodman's use of link-tracing is
motivated by improvements in efficiency allowed by over-sampling
relations most likely involved in the structures he is studying.

More recently, the term ``snowball sampling" has been taken to
refer to a convenience sampling mechanism with motivation more
like that of Trow: collecting a sample from a population in which
a standard sampling approach is either impossible or
prohibitively expensive, for the purpose of studying
characteristics of individuals in the population 
\citet[e.g., ]{biernackiwaldorf1981}.  Such settings are often {\it hard-to-reach}
populations, characterized by the lack of a serviceable sampling
frame.  In such cases, an initial probability sample is either
impossible or impractical, such that the initial sample is drawn
by a convenience mechanism, dooming the full sample to
non-probability sample status.  In many such hard-to-reach
populations, link-tracing sampling is an effective means of
collecting data on population members.  For this reason, this
latter non-probabilistic usage of ``snowball sampling" is most
common in practice, although less common in the statistical
literature, which favors the probabilistic formulations.  Note
that it is possible for the seeds in RDS to be chosen randomly
even in applications to hard-to-reach populations.  For example,
they could be selected based on a spatial sampling frame.

The tension between these two uses of snowball sampling is
highlighted in \citet{thompson2002}, a definitive textbook, (p.  183):
``The term `snowball sampling' has been applied to two types of
procedures related to network sampling.  In one type ..., a few
identified members of a rare population are asked to identify
other members of the population, those so identified are asked to
identify others, and so, for the purpose of obtaining a
nonprobability sample or for constructing a frame from which to
sample.  In the other type (Goodman 1961), individuals in the
sample are asked to identify other individuals, for a fixed
number of stages, for the purpose of estimating the number of
`mutual relationships' or `social circles' in the population."
Other definitions of ``snowball sampling" are consistent with
this duality in usage \citep[p. 59]{snijders1992}.

Respondent-driven sampling \citep[RDS, introduced by Heckathorn and
colleagues, e.g.][]{heck97} is a newer variant of
link-tracing network sampling, which brings to a head the tension
between these two usages.  This is because RDS is a practical
sampling method in hard-to-reach populations, beginning with a
convenience sample, but aims to approximate a probability sample
over time.

RDS is not a variant of either usage of snowball sampling, nor is
the reverse true.  Because of the confusion surrounding this
term, in \citet{gilehansocmeth10} we prefer, and use throughout
that paper, the more precise broad category ``link-tracing
sampling" while paying homage to the intellectual descent of the
methods from snowball sampling.

It is precisely the tension between the two usages of snowball
sampling that makes RDS a fruitful area for ongoing research.
RDS pairs the practical implementation of a convenience sample
with the hope of recovering ``something like" a probability
sample.  \citet{gile08} and \citet{gilehansocmeth10} are the first
works to systematically evaluate the statistical properties of
current estimators based on RDS data.  \citet{gilejasa11} proposes a new
estimator that adjusts for the bias introduced by the
with-replacement assumption of these estimators. 
It is also sometimes possible to adjust for a convenience sample
of seeds.  For example, \citet{gilehan11MA} extend the estimator
of \citet{gilejasa11} to correct for the bias introduced by seed
selection in the presence of homophily.

The issue here, then, is to recognize the different uses of the
term ``snowball sampling".  A good solution is for scientists to
be as clear as possible in defining the meaning of terms upon
first use in each manuscript.  There is enough confusion in the
various literatures to make this good practice.

\bibliographystyle{plainnat}
\bibliography{onsnowballsampling}

\end{document}